# Ear Wearable (Earable) User Authentication via Acoustic Toothprint

ZI WANG, Florida State University, USA
YILI REN, Florida State University, USA
YINGYING CHEN, Rutgers University, USA
JIE YANG, Florida State University, USA

Earables (ear wearable) are rapidly emerging as a new platform encompassing a diverse range of personal applications. The traditional authentication methods hence become less applicable and inconvenient for earables due to their limited input interface. Nevertheless, earables often feature rich around-the-head sensing capability that can be leveraged to capture new types of biometrics. In this work, we propose ToothSonic that leverages the toothprint-induced sonic effect produced by a user performing teeth gestures for earable authentication. In particular, we design representative teeth gestures that can produce effective sonic waves carrying the information of the toothprint. To reliably capture the acoustic toothprint, it leverages the occlusion effect of the ear canal and the inward-facing microphone of the earables. It then extracts multi-level acoustic features to reflect the intrinsic toothprint information for authentication. The key advantages of ToothSonic are that it is suitable for earables and is resistant to various spoofing attacks as the acoustic toothprint is captured via the user's private teeth-ear channel that modulates and encrypts the sonic waves. Our experiment studies with 25 participants show that ToothSonic achieves up to 95% accuracy with only one of the users' tooth gestures.

## 1 INTRODUCTION

Earables are rapidly emerging as a new platform encompassing a broad range of personal applications due to their rich around-the-head sensing capability [10, 43]. A recent report shows that earables like Apple's AirPods have much stronger and broader demand than smartwatches and fitness trackers, and are the driving force of the wearable market growth [81]. Current projections indicate the market for earables will reach over 5 billion dollars by 2030, and they are becoming more intelligent [23]. There also have been increasing research efforts to leverage earables for speech recognition [30, 43], augmented reality [10, 80], motion and activity tracking [44, 50], healthcare monitoring [7, 49], and context information extraction [10, 29].

Authors' addresses: Zi Wang, Florida State University, 1017 Academic Way, Tallahassee, FL, 32306, USA, ziwang@cs.fsu.edu; Yili Ren, Florida State University, 1017 Academic Way, Tallahassee, FL, 32306, USA, ren@cs.fsu.edu; Yingying Chen, Rutgers University, 96 Frelinghuysen Rd, Piscataway, NJ, 08854, USA, yingche@scarletmail.rutgers.edu; Jie Yang, Florida State University, 1017 Academic Way, Tallahassee, FL, 32306, USA, jie.yang@cs.fsu.edu.

While earables show considerable promise, they also raise new questions in terms of security. This is because many services offered by earables depend on the confidential and personal information they capture, process and transmit [43]. Moreover, earables have also considerable promise as tokens that mediate access to online accounts and a diverse set of devices in smart environments [10]. It is thus critical to developing secure authentication for earables to prevent unauthorized access to security-sensitive data and services.

However, directly adapting traditional authentication from other wearables or mobiles can be challenging. This is simply because earables lack a suitable input interface to support rapid and reliable entry of passwords or most of the traditional biometrics [10]. Voice-based authentication is convenient for earables but has been proven vulnerable to voice spoofing attacks [27, 82]. Despite the issue, earables also provide novel opportunities to improve or redesign approaches to authentication due to their rich around-the-head sensing capability [75]. For example, recent work utilizes earable to sense ear canal [12] and its deformation [76] for authentication. However, actively emitting acoustic sound to probe the ear canal could be intrusive for those who are sensitive to high-frequency sound.

In this work, we propose ToothSonic, a secure authentication system that leverages the toothprint-induced sonic effect produced by a user performing teeth gestures for earable authentication. In particular, when teeth slide or strike against each other, part of their mechanical energy is transformed into a sonic wave. The harmonics of the friction- and collision- excited sonic waves are dependent on the teeth composition, the dental geometry, and the surface characteristics of each tooth [3]. The key insight is that the sonic waves produced from a teeth gesture (either slide or tap) carry the information of the toothprint. As every individual has a unique toothprint just like our fingerprint, two users perform the same teeth gesture will result in distinct toothprint-induced sonic waves, which could be sensed by the earables for user authentication, as shown in Fig. 1. Compared with traditional biometrics, it has several advantages.

**Anti-Spoofing.** The friction- and collision- excited sonic waves are dependent on the toothprint, which is hidden in the mouth and skull. It is thus more resilient to spoofing attacks compared with traditional biometrics (e.g., fingerprint, face, and voice) that could be easily stolen by adversaries [9]. In addition, the sonic waves travel through the head tissues and skull channel, which hold the individual uniqueness acting as a private secure channel that modulates and encrypts the sonic waves. ToothSonic is thus resistant to sophisticated adversaries who can acquire the victim's toothprint, for example, via the dentist.

**Wide Acceptability.** ToothSonic provides eye-free and hands-free authentication when hands and eyes are occupied (e.g., carrying or driving), whereas most traditional biometric approaches require explicit user operation, such as pressing the fingertip on a reader or posing the face or eyes to the camera. It is also more socially acceptable than voice-based authentication in public places (e.g., offices and libraries) as the sonic waves of teeth gestures are much less perceptible and unobtrusive to others, which also protects user privacy as oppose to audible voice password or paraphrase.

**Implicit Authentication.** ToothSonic can be further exploited as an implicit authentication method when teeth gestures are used as a hands-free computer interface, for example potentially in "Switch Access" services, and for people with motor impairments [51].

In our work, we design a set of representative teeth gestures based on the factors that impact the toothprint-induced sonic waves. Specifically, we choose six sliding gestures and four tapping gestures to represent multi-level characteristics of teeth as well as to balance the easy to perform. These gestures can produce effective sonic waves that carry the information of the toothprint. As the sonic waves propagate through the head tissues and skull to the ear canal, they will be significantly attenuated. To reliably capture the attenuated sonic waves, we choose the inward-facing microphone among various embedded sensors on earables. Utilizing the inward-facing microphone has one key advantage that the earbud and the ear canal form the occlusion effect, which boosts the sonic waves, especially for the low-frequency part that carries effective information of toothprint.

The sensed sonic waves are then going through the pre-processing to remove noise and to segment the data of each gesture. Then, our system extracts multi-dimensional acoustic features that correspond to the multi-level acoustic toothprint, and compares these features against the user enrolled profile to perform authentication. To evaluate ToothSoinc, we conduct experiments with 25 subjects under various scenarios with different teeth gestures. The contributions of our work are summarized as follows:

- We explore the toothprint-induced sonic waves produced by teeth gestures for user authentication. It is designed for earables and has unique advantages when compared to traditional biometrics.
- We investigate representative teeth gestures that produce effective sonic waves carrying the toothprint information. We also leverage the occlusion effect to reliably capture the sonic waves, and extract multi-level features to represent intrinsic toothprint information.
- We validate the ToothSonic through both experiments and user surveys. Results show that it achieves 97% accuracy with one of the users' preferred gestures.

## 2 RELATED WORKS

### 2.1 Biometric Based Authentication

**Traditional Biometric Based Authentication.** The commonly used biometrics rely on physiological and behavioral characteristics. One category of biometric used at large rifeness is based on human physiological characteristics, such as fingerprint, palmprints, face. The fingerprint-based has been widely used in many applications and quickly integrated into mobiles, wearables, and IoT devices in the recent past [13, 39]. It has also been widely used in forensic medicine, government, and municipal administration [13, 39]. However, the researcher found that malicious attacks such as fingerprint obfuscation and impersonation could make fingerprint-based authentication vulnerable [40, 41]. It is also worth noticing that fingerprint is easy to be assailable by bolted or removed fingertips, or even fooled by the mimic attack, which replaces part of the fingertip's skin. For instance, using materials such as silicon, artificial fingerprints could be acquired from latent fingerprints left on objects [1] to deceive the recognition systems [61]. Palmprints work similar to the fingerprint-based authentication systems [24, 26], however, they are also vulnerable to spoofing attacks. Meanwhile, face authentication is another popular biometric and are supported on most smartphones now [4, 72]. Newly released smartphones such as those manufactured by Apple, Samsung have all included face authentication for unlocking the phone and other purposes [4, 72]. However, recent work shows these biometrics are suspectable to spoofing attacks [9, 45]. In the meantime, facial authentication is also vulnerable to face morphing attacks [14, 56]. In the recent COVID-19 pandemic, face-based authentication has also been proven to not work well when the user is wearing a mask.

Some other popular ones such as voice [84], gait [52, 53], signature [54, 55], vital signs [37, 38], figure gestures [67, 68], activities [69, 74], and locations [86, 87] belong to behavioral characteristics category. Due to the wide acceptance of smart speaker in the IoT environment and on mobiles, voice-based user authentication has become most beloved and integrated with many devices. However, recent studies found that the voice biometric is vulnerable to spoofing attacks [27, 82, 83], such as replay attacks [11, 18] and speech-synthesize attacks [78].

**Functional Biometrics Based Authentication.** Many recent studies focus on studying the Functional Biometric [36]. Functional Biometrics utilize the human body as a transfer function and generate a unique response to a stimulus that is used as biometrics for user authentication. For instance, SkullConduct [60] utilizes a bone-conducted acoustic stimulus and captures the characteristic frequency response of the human, which serves as a robust biometric. More recent proposed biometrics includes ear-canal and its deformation. For instance, Arakawa et al. [2] propose to utilize the earphones to capture the static ear canal geometry, it extracts the Mel-frequency cepstral coefficients (MFCCs) features of the reflected acoustic signals from the ear canal to distinguish different users. Similarly, EarEcho [16] captures the acoustic characteristics of the static geometry of the ear canal for user authentication. Meanwhile, Wang et al. [76] use the in-ear wearables to captures the ear canal deformation for continuous authentication. However, system such as EarEcho requires emitting sound to probe the ear canal, which could be intrusive for those who are sensitive to the probe sound.

**Implicit authentication.** In addition, both behavior biometric and functional biometric could be naturally combined with implicit authentication [25, 59, 60]. Implicit authentication is the approach that allows system to authenticate their user by the observations of the user's behavior for authentication. Different from the explicit authentication which require user input, the implicit authentication enables the ability to authenticate users based on actions they would perform anyway. Implicit authentication could enable some emerging applications such as Augment Reality and Virtual Reality.

### 2.2 Tooth Based Identification and Computing

**Tooth-Based Identification.** Tooth has high resistance against high temperature and corrosion due to the enamel around it. Different from other biometrics such as face and fingerprint, which may be easily damaged, tooth plays a core role in identification the deceased individuals and resolving violent crimes based on both tooth dental works and tooth DNA [66]. Compared with traditional biometrics, toothprint is more resistant to spoofing attacks and has been used for individual identification. For example, some researchers compare X-ray film and the dental records of users to identify if they are the same individuals [20, 34]. Wang *et al.* [73] proposed a dental impression-based user identification that extracts the structure features from dental impression images for identification. Although these systems could achieve good accuracy on user identification, they require dedicated devices and the processes are expensive and time-consuming. Recent studies explore mobile devices to sense the toothprint. For instance, SmiltAuth [28] utilizes smartphone to take photos of the dental edge for user authentication, whereas BiLock [85] uses mobile devices to capture the sound of dental occlusion for authentication. These methods, however, are not suitable for earables and measure only one teeth gesture, i.e. occlusion tapping. Meanwhile, it also requires the user to take extra efforts, such as put their hand near their mouth to capture the sound of tooth click (i.e. smartwatch on wrist or smartphone holding in hand). Compared with these works, our work focuses on modeling and understanding macro and micro biometrics of human toothprint-induced sonic waves. Our system seeks to obtain multiple teeth gestures carried enriching biometric represents multi-level toothprint biometric. In addition, our system also leverages the hidden channels of the face and skull and the occlusion effect to enhance the performance of our system and prevent our system against attacks such as side-channel attacks and mimic attacks.

**Tooth-Based HCI.** There have been research efforts to sense teeth gestures for human-computer interaction, including oral sensor-based and ear sensor-based approaches. Oral sensors-based systems leverage invasive sensors mounted inside the mouth to sense the tooth and tongue related gestures. For example, Clench Interface [79] adopted a pressure sensor between human teeth to enable clenching as an interface. ChewIt. [15] proposed in-mouth edible devices to sense teeth gestures. Ear sensors-based systems utilize various in-ear sensors to identify oral activities. For instance, EarSense [51] uses the earable to capture the teeth sound for teeth gestures recognition. TeethTap [65] utilizes an IMU sensor and a microphone on earpieces to identify teeth gestures. Byte.it [71] uses the motion sensor on earables to recognize various teeth clicks. Our proposed ToothSonic can be integrated with these systems to enable implicit and continuous user authentication.

## 3 PRELIMINARIES

### 3.1 Human Teeth Biometric

Humans are diphyodonts and thus have two sets of teeth with the first set being called deciduous teeth that contains 20 teeth, which will be replaced by the second set of permanent teeth [62]. Permanent teeth normally have 32 teeth, where 16 are in the maxillary (upper teeth) and 16 in the mandibular (lower teeth). The teeth are symmetrical on both upper/lower and left/right, thus can be divided into four quadrants, as shown in Fig. 2. The dental formula is 2.1.2.3, which means we have 2 incisors, 1 canine, 2 premolars, and 3 molars in each quadrant.

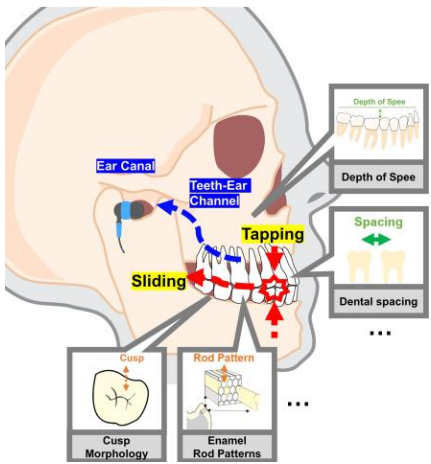
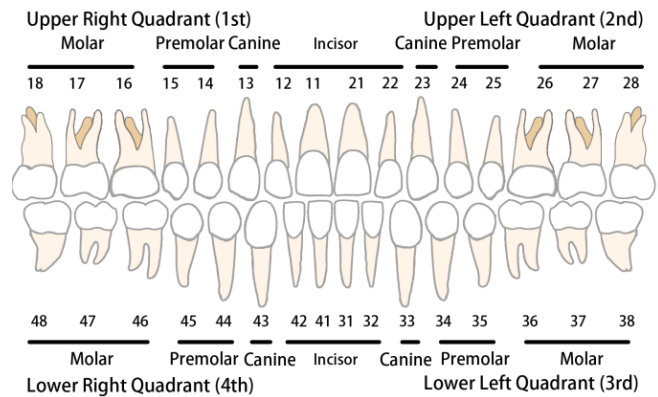

Fig. 1. The core idea of ToothSonic.

Fig. 2. Human Dentition System.

Toothprint has been used in forensic identification because dental biometrics show uniqueness across individuals, as well as the tooth show high resistance against high temperature and corrosion [17]. In particular, the teeth biometric could be summarized into two categories: innate and nurtured characteristics. The innate ones mainly include *(i) dental geometry*, *(ii) single tooth*, and *(iii) tooth anomalies*, whereas nurtured ones refer to **nurtured dental works**.

For *dental geometry*, the information of the dental arch, the depth of spee, dental spacing, and occlusion classes shows unique characteristics across individuals. As shown in Fig. 4(a), the dental arch presents individual diversity. For example, the depth and width of the mandibular arch can vary from 39.5mm to 51.3mm and from 53.3 to 63.8mm for different persons, respectively. The maxillary arch has similar diversities as well [6]. Moreover, the depth of spee shows uniqueness as it measures the distance between the deepest cusp tip and a flat plane of the top of the mandibular dental cast, as shown in the upper right in Fig. 1. According to dental records over 26 years, the depth of spee remains unchanged for adults with a mean of -3.33mm and a max of -0.47mm [42]. Similarity, dental spacing, and occlusion classes of teeth present distinct features for each individual [5].

The *single tooth* characteristics include cusp morphology, enamel rod patterns, and enamel thickness [19], as shown in the bottom two figures of Fig. 1 and Fig. 4(b). For example, the number of cusps on the crown of the molar could range from 4 to 7 for different individuals [48]. The enamel rod patterns refer to groups of enamel rods on the tooth surface. They have been used as one type of teeth print for person identification[47]. Moreover, the enamel thickness could range from 0 to 1617 microns with a standard deviation of 221 microns [46]. And, female teeth have proportionately more enamel and male teeth have more dentine [63].

Moreover, *teeth anomalies* [33, 70] such as twisted/tilted teeth, teeth rotations/transpositions, missing/extra teeth, supernumerary cusps also exists. The other category of dental diversity is *nurtured dental works*, such as orthodontic treatment, dental fillings, dental implants, tooth extractions, surface structure/root configuration, dental veneers, dental crowns, and so on. Tradition forensic dentists usually utilize these dental works for identification by comparing the target with their dental work record. Our system benefits from both teeth anomalies and nurtured dental works as they both introduce personal uniqueness.

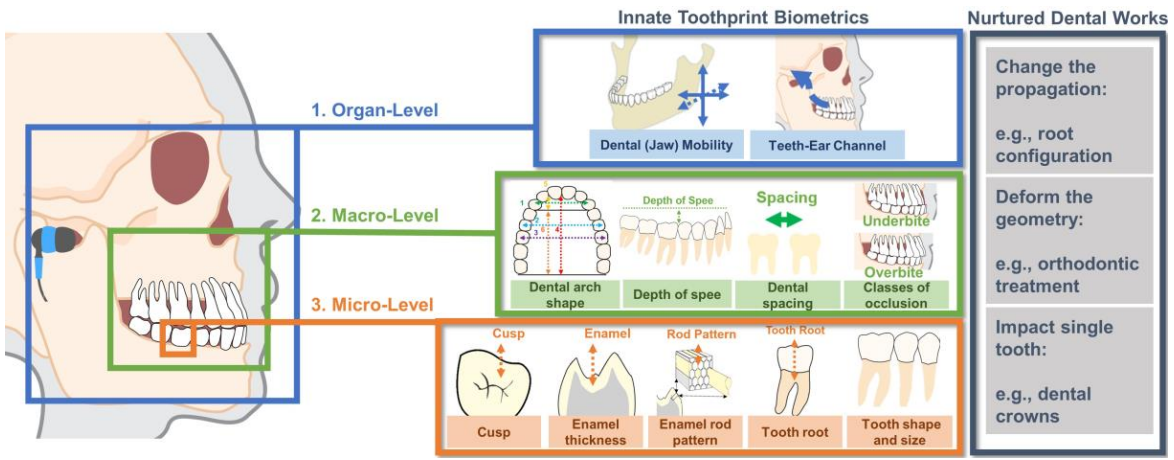

Fig. 3. Multi-level impact factors to acoustic toothprint.

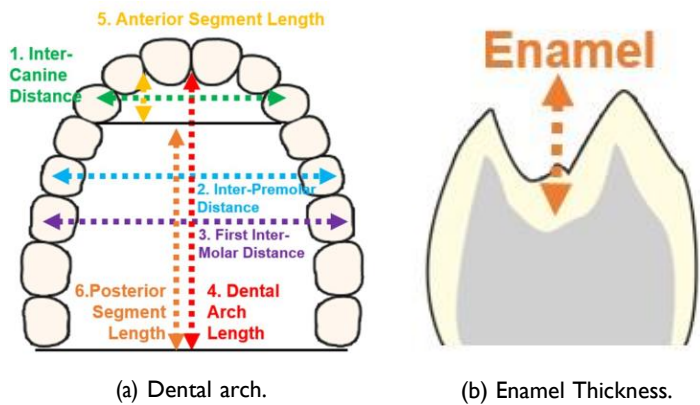

(a) Dental arch.      (b) Enamel Thickness.

Fig. 4. Examples of tooth biometrics.

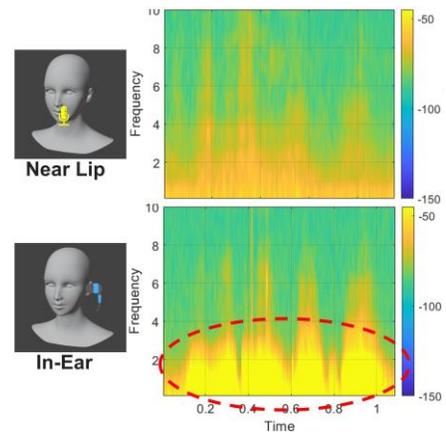

Fig. 5. Occlusion effect of the ear canal.

## 3.2 Impact Factors to Acoustic Toothprint

When the mandibular (lower tooth) and maxilla (upper tooth) are sliding or tapping against each other, part of their mechanical energy is transformed into sonic waves. The harmonics of such sonic waves are dependent on the teeth characteristics, and thus carrying the information of the toothprint. In this section, we analyze the impact factors of the toothprint-induced sonic waves when performing teeth gestures. It serves as the basic principle to design appropriate teeth gestures that could produce information-rich sonic waves for characterizing the toothprint. In particular, we propose a taxonomy that highlights the impact factors with three levels: ***organ-level***, ***macro-level*** and ***micro-level*** factors, as shown in Fig. 3.

The *organ-level* impact factors include the dental mobility and the propagation channel. Dental mobility refers to the control of the mandibular, whereas the propagation channel is the pathway that the sonic waves transmitted from the teeth to the ear canal. Dental mobility has 3 axes of freedom, i.e., up/down, left/right, and in/out [51]. It thus controls the degree of teeth sliding and the scale of mouth open for tooth tapping. It further impacts how many teeth participate in the gestures that produce toothprint-induced sonic waves.

The *macro-level* mainly includes the factors related to the dental geometry, dental spacing, classes of occlusion, and the depth of spee. As shown in Fig. 4(a), the dental arch will determine how many teeth participate in tapping and sliding. If the upper and lower teeth are highly coordinated, then most teeth will participant in the teeth gestures, and thus generating information-rich sonic waves carrying more information. Otherwise, less informative sonic waves will be generated. Moreover, the depth of spee determines if premolar or molar can interact with each other. Thus, if the depth of spee is too big, the premolar would not generate any sonic waves when performing gestures. In addition, the tapping (between upper and lower teeth) and clicking (between two teeth) produce the sonorant sonic waves. Furthermore, the class of occlusion determines the contact surface of each teeth gesture, thus affecting the sound waves generation. And there are three major occlusion classes (i.e., *normal bit*, *overbit*, and *underbit*), which are defined based on the amount of vertical overlap between the upper and lower front teeth.

The *micro-level* includes the characteristics of each tooth, such as cusp morphology, enamel thickness, enamel rod patterns, tooth root, the shape and size of each tooth, and nurtured dental works [19]. The cusp morphology tells us how many cusps are participating in the generation of sonic waves and whether they produced a high-pitched wave. Moreover, different thicknesses of enamel thickness produce different frequencies and tones when tapping. In addition, the rod patterns impact the friction of sliding and contribute to the fricative sonic waves. The tooth root and nurtured dental works are part of the teeth composition, which modulates the sound waves.

### 3.3 Sensing Acoustic Toothprint

One unique advantage of using earable to sense the toothprint-induced sonic waves is that the sensed sonic waves are modulated by the teeth-to-ear propagation channel of the user. As this channel is user-dependent, it is unlikely, if not possible, for an adversary to obtain such a unique private channel to simulate the sensed sonic waves at earable. Thus, the toothprint-induced sonic waves travel through the head tissues and skull, which act as a private secure channel that modulates and encrypts the sonic waves.

While the head tissues and skull channel secure the sonic waves, they also significantly attenuate the signals. Among various embedded sensors available on earables, such as IMU sensors, inward- and outward- facing microphones, we choose the inward-facing microphone to capture the attenuated sonic waves. The reason to choose the inward-facing microphone is that the *occlusion effect* could be leveraged to boosts the low-frequency sonic waves that carry the information of toothprint. Specifically, when the ear canal is blocked by the earbud, the sonic waves will be rebounded to the eardrum, thus amplifying the low-frequency part of the sonic wave [58, 64]. Fig. 5 shows the occlusion effect captured by the inward-facing microphone when comparing to using the external microphone to record sonic waves. We can observe that low-frequency components in the red circle are amplified due to the occlusion effect.

## 4 SYSTEM AND ATTACK MODELS

Our system requires the earables equipped with an inward-facing microphone to capture the toothprint-induced sonic waves. The authentication could be triggered on-demand, or can be used as an implicit authentication method when teeth gestures are used as a hands-free computer interface. To authenticate a user, the user is required to wearable the earable and to perform one or more teeth gestures, and the corresponding profile of the user's acoustic toothprint should be enrolled ahead offline.

In this work, we primarily consider three types of spoofing attacks: *mimic attacks*, *advanced mimic attacks*, and *replay attacks*. For mimic attacks, adversaries wear the victim's earable and perform the same gestures as the victim did in an attempt to pass the authentication. For advanced mimic attacks, we assume sophisticated adversaries can acquire the victim's toothprint. They can precisely rebuild the teeth of the victim (e.g., using 3D printing) and then use it to perform teeth gestures for spoofing attacks. We also consider replay attacks, where the adversaries first try to eavesdrop on the sonic wave of the victim's toothprint, for example, by recording the sound in close proximity of the victim. Then, they replay the recorded one to the authentication system to conduct spoofing attacks.

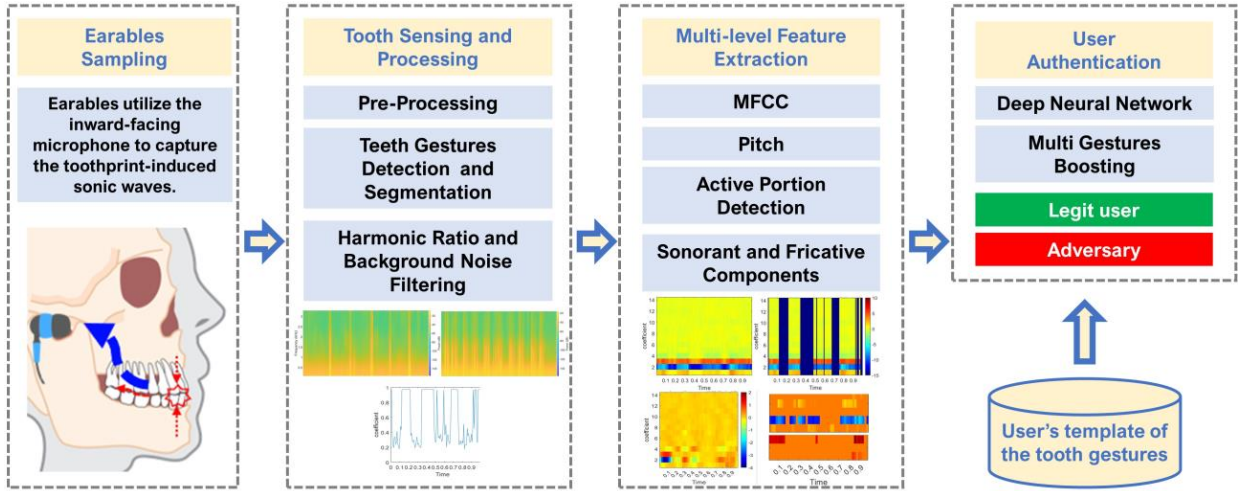

Fig. 6. ToothSonic system flow.

Our proposed ToothSonic can effectively defeat these three types of attacks. First, the sonic waves generated from the teeth gestures are dependent on the toothprint, which is unique to each individual and is hidden in the mouth and skull. It is thus resilient to simple mimic attacks. Second, the toothprint-induced sonic waves are captured via the user's private teeth-ear channel. Our system thus is resistant to advanced mimic and replay attacks as the user's private teeth-ear channel secures the sonic waves, which are unlikely uncovered by adversaries.

## 5 SYSTEM DESIGN

### 5.1 System Overview

The basic idea of our system is to leverage the toothprint-induced sonic waves produced by a user performing teeth gestures for secure earable authentication. As illustrated in Fig. 6, the system takes as input time-series sonic waves that are sensed at the earable when a user performing teeth gestures. The teeth-to-ear channel of the user serves as a private secure channel that modulates the transmitted sonic waves, which are then preprocessed to detect and segment the teeth gestures and also to remove noises and outliers via harmonic ratio and background noise filters.

The core of our system, ToothSonic, is the *Teeth Gesture Design* and *Feature Extraction*, which will be elaborated on in sections 3.2 and 3.3. For gesture design, we study a set of representative teeth gestures based on both the factors that impact the toothprint-induced sonic waves and the usability of the gestures. For feature extraction, our system extracts multi-dimensional acoustic features, such as MFCC, Pitch, log spectrum, sonorant and fricative components, to represent multi-level information of the toothprint. Next, our system authenticates a user by inputting the extracted features into a deep neural network. ToothSonic can authenticate a user based on either a single teeth gesture or a combination of multiple teeth gestures.

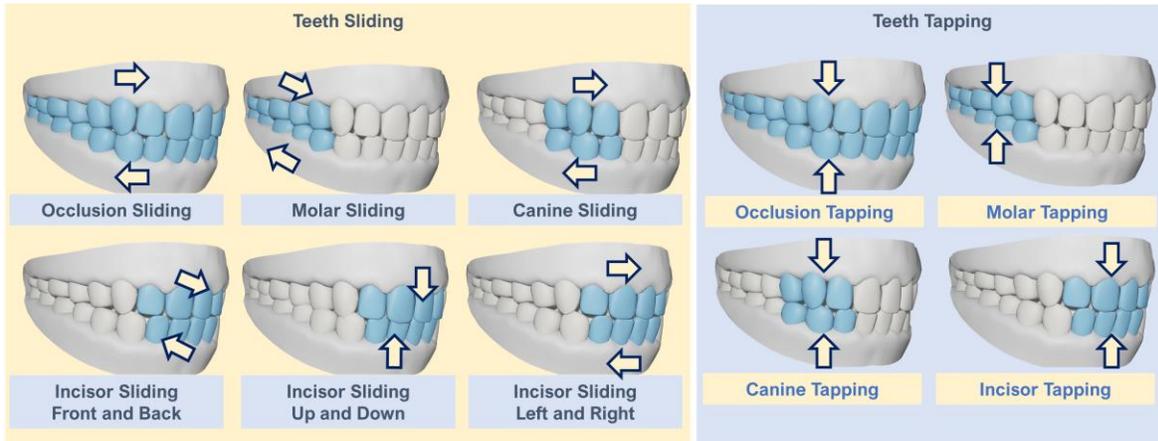

Fig. 7. Illustration of ten advanced teeth gestures.

| No. | Biometric Factors<br>Tooth Gestures | (1) Organ-level | | | | (2) Macro-level | | | | | | | (3) Micro-level | | | |
|---|---|---|---|---|---|---|---|---|---|---|---|---|---|---|---|---|
| | | 1. Dental Mobility F/B | 2. Dental Mobility U/D | 3. Dental Mobility L/R | 4. Propagation Channel | 5. Dental arch shape | 6. Depth of spee | 7. Occlusion classes | 8. Dental spacing | 9. Incisor shape and size | 10. Canine shape and size | 11. Molar shape and size | 12. Cusp | 13. Enamel thickness | 14. Enamel rod patterns | 15. Tooth root |
| 1 | Occlusion Sliding | | √ | | √ | √ | √ | √ | √ | √ | √ | √ | √ | √ | √ | √ |
| 2 | Molar Sliding | √ | | | √ | | √ | | √ | | | √ | √ | √ | √ | √ |
| 3 | Canine Sliding | √ | | | √ | | | | √ | √ | √ | | | √ | √ | √ |
| 4 | Incisor Sliding F/B | √ | | | √ | | | √ | √ | √ | √ | | | √ | √ | √ |
| 5 | Incisor Sliding U/D | | √ | | √ | | | √ | √ | √ | √ | | | √ | √ | √ |
| 6 | Incisor Sliding L/R | | | √ | √ | | | √ | √ | √ | √ | | | √ | √ | √ |
| 7 | Occlusion Tapping | | √ | | √ | √ | √ | √ | | √ | √ | √ | √ | √ | | √ |
| 8 | Molar Tapping | | √ | | √ | | √ | | | | | √ | √ | √ | | √ |
| 9 | Canine Tapping | | √ | | √ | | √ | | | √ | √ | √ | | √ | | √ |
| 10 | Incisor Tapping | | √ | | √ | | | √ | | √ | √ | | | √ | | √ |

Fig. 8. Teeth gestures and the toothprint information carried by the sonic waves.

## 5.2 Teeth Gesture Design

When designing teeth gestures, we mainly consider two factors: what toothprint information could be possibly carried by the sonic waves produced from the gestures, and how easily could a user perform such gestures. Based on the multiple impact factors we discussed in section 2.2, we design several candidates for both sliding and tapping gestures. Then, we invite the participants to perform these candidate gestures, and make a survey to find out the gestures that the majority of the participates can comfortably execute. Eventually, we advance 10 teeth gestures for further study.

Fig. 7 illustrates how these gestures are performed. The ten teeth gestures include six sliding gestures and four tapping gestures. The sliding gestures contain *occlusion sliding*, *molar sliding*, *canine sliding*, *incisor sliding front/back*, *incisor sliding up/down*, and *incisor sliding left/right*. And the tapping gestures are *occlusion tapping*, *molar tapping*, *canine tapping*, and *incisor tapping*.

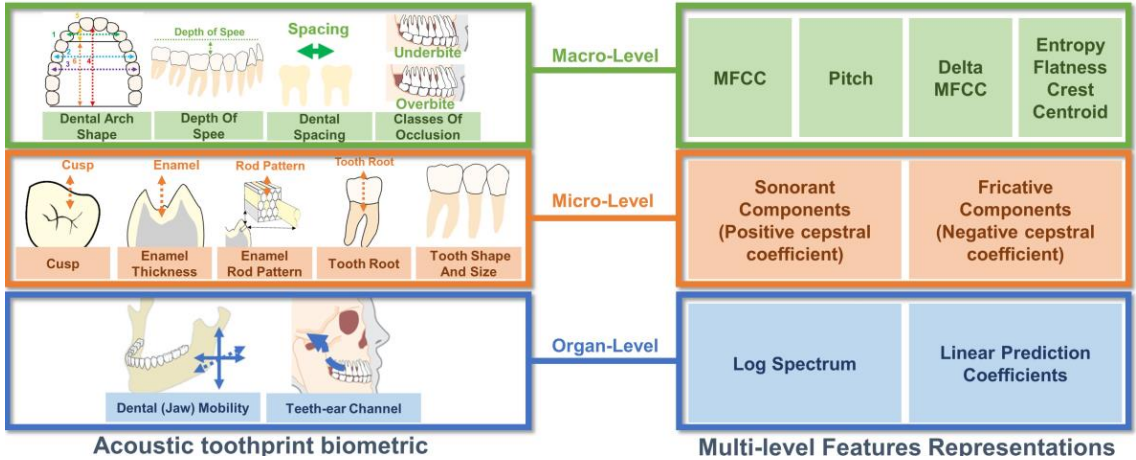

Fig. 9. Multi-level features and the corresponding intrinsic toothprint.

In addition, Fig. 8 describes the possible toothprint information that could be carried by the sonic waves from each gesture. The first six rows reveal the details of 6 sliding gestures. At high-level, sliding requires the mandibular to move in one direction and then move back. For instance, sliding front and back (or left and right) will reflect the dental mobility in the F/B (or L/R) direction. The sonic waves generated from sliding gestures contain the information of the enamel rod patterns and the enamel thickness, which are represented as the fricative components of the sonic waves. Moreover, the sonorant portions of the sonic waves carry the information of the dental spacing when sliding. In addition, in *molar sliding*, the cusps on the molars will grind, thus producing the high-frequency components of sonic waves. However, if the depth of spee is deep, then the upper and lower pre-molar have little chance to interact with each other, hence resulting in fewer high-frequency components. Furthermore, if the upper and lower teeth are highly coordinated, most teeth will be involved in the teeth gestures, and thus generating information-rich sonic waves.

The last four rows in Fig. 8 illustrate these four tapping gestures. The sonic waves generated by tapping gestures reveal the mobility on the Z-axis, i.e., up and down. And the sonic waves of tapping gestures could disclose the enamel thickness as the enamel dominates the contacting area when tapping. Moreover, the sonic waves of different tapping gestures reflect the features of different teeth involved. For example, the sonic waves of *molar tapping* reveal toothprint of molars only. Similar to sliding, the sonic waves of the tapping also carry the information of the upper and lower arch. Furthermore, the sonic waves induced by *occlusion tapping* reflect the information of occlusion classes, i.e., *normal bit, overbit or underbit*. For example, for *overbite*, the upper inner surface is detached from the lower teeth, thus producing no sonic waves, whereas for *normal bite*, the upper inner surface contributes a lot to the sonic waves. More details could be found in Fig. 8.

### 5.3 Multi-level Feature Extraction

Our system extracts multi-dimensional acoustic features to represent multi-level information of the toothprint. In particular, these multi-dimensional features are distributed into three categories, as shown in Fig. 9.

**Macro-level features.** We first extract the Mel frequency cepstral coefficients (MFCC) and Pitch to reveal the macro-level characteristics of toothprint [22, 77]. In particular, the tapping and sliding segmentation are first converted into Mel-frequency cepstrum and then go through pre-define filter banks (i.e., the different Mel filters in Mel-scale). After that, logarithmic compression and discrete cosine transform are executed. Then, 14-dimensional

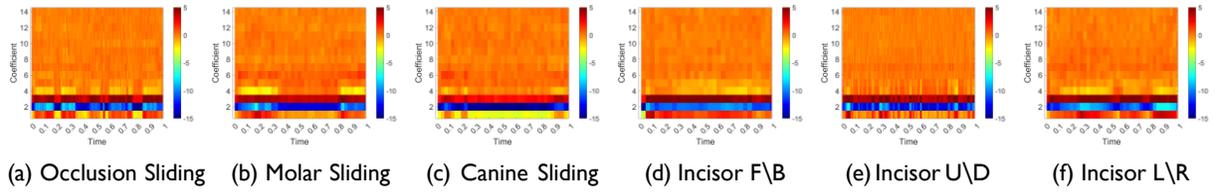

Fig. 10. MFCC features for the same subject with 6 different sliding gestures.

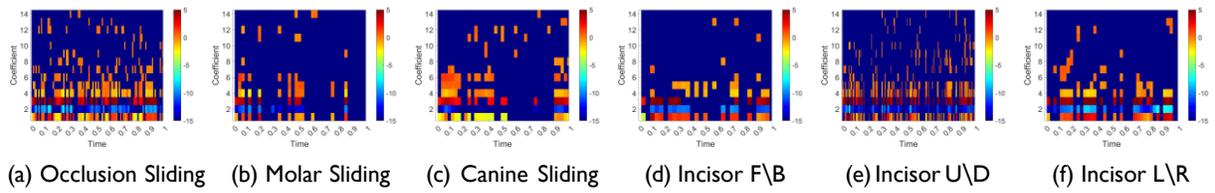

Fig. 11. Active portion features of 6 different sliding gestures.

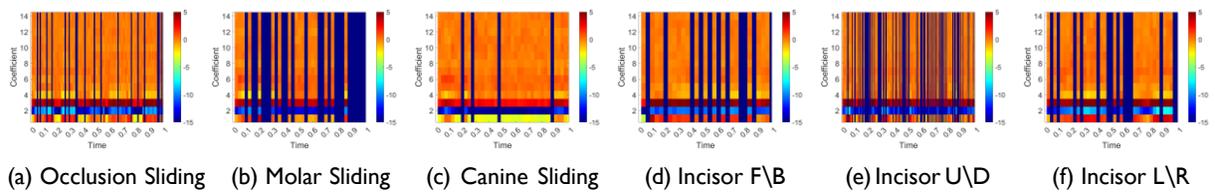

Fig. 12. Harmonic portion extraction of 6 different sliding gestures.

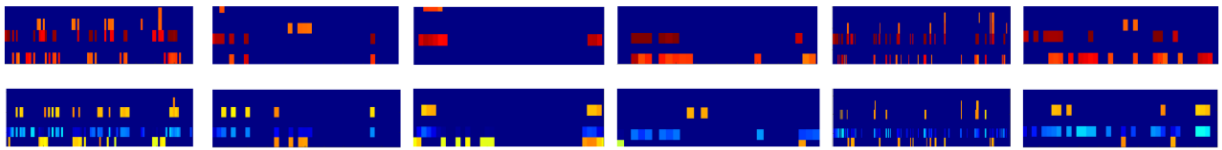

Fig. 13. Sonorant (upper) and fricative (lower) components of sonic waves across six sliding gestures.

MFCCs are extracted as shown in the Y-axis of Fig. 10, which illustrates the MFCC features extracted for six sliding gestures from the same subject. It could be observed that the first 6 dimensions dominant the information in the received sonic waves. Moreover, different gestures of the same subject show distinct patterns.

Moreover, to better locate where the sonic waves generated most frequently. We find out the active portion (i.e., the time when teeth hit each other) of the teeth gesture based on the delta MFCC [21]. We detect the active portion when the corresponding Delta-MFCC is bigger than a pre-defined threshold. We then focus on the features extracted from these active portions. As shown in Fig. 11, the features of the active portions have high entropy, thus carrying the majority of the toothprint information. In addition, at the macro-level, spectral features are also acquired for tapping gestures including entropy, flatness, crest, and centroid.

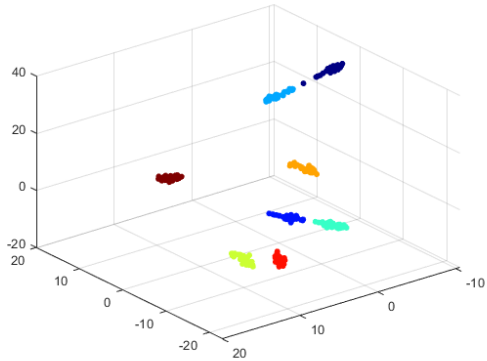

Fig. 14. Features dimensionality reduction across different subjects using t-SNE.

**Micro-level features.** We extract the sonorant and fricative components of the sonic waves to reveal the micro-level toothprint biometric. Specifically, the fricative components in the sonic waves are mainly produced by enamel rod patterns on the teeth when sliding, whereas the sonorant portion of sonic waves is dominated by the cusp and enamel thickness when performing gestures. To extract the sonorant and fricative components, we check the value of the cepstral coefficient. In particular, for those with a positive value, the sonic wave tends to be sonorant components, whereas for the negative values, the corresponding sonic waves are contributed more by the fricative ones. The extracted sonorant and fricative components of six different sliding gestures are illustrated in Fig. 13. They also reveal other important toothprint information, such as dental spacing, as the vertical blue lines portions shown in the upper part of Fig. 12.

**Organ-level features.** We extract the features of log spectrum and linear prediction coefficients to capture the hidden channel of the head and skull.

To illustrate how these features could be useful to differentiate different users, we utilize t-SNE to reduce the feature space to three major dimensions and plot the Fig. 14. It shows the three major dimension features of the *occlusion sliding* gesture across 8 different subjects, with each subject tried 40 times. It could be observed that the gestures performed by the same user are clustered together, whereas the gestures for different users are separated into different clusters. It shows a high intra-class correlation and large inter-class distances, thus indicating the effectiveness of the extracted features.

### 5.4 System Implementation

**Pre-Processing.** The toothprint-induced sonic wave is subtle and easily affected by interferences. By analyzing the spectrum of our received sonic waves, we find that the major component of toothprint-induced sonic waves is located under 8kHz. Meanwhile, there could be some human body vibrations that interfere with the sonic waves, such as moving arms and legs. Thus we set a filter band as [20Hz, 8000Hz] to filter out-of-band background interferences.

**Teeth Gesture Segmentation.** After the noise filtering, our next step is to segment the sonic waves into each gesture. In particular, we utilize the Hidden Markov model (HMM) as the primary way of sonic wave segmentation [32, 57]. It is done by combining the methods of HMM based statistical classification and probabilistic rule based method. It first computes the probability distribution of the input pieces of the sonic wave and compares with the components to search for the highest probability. Then only the weighted component with

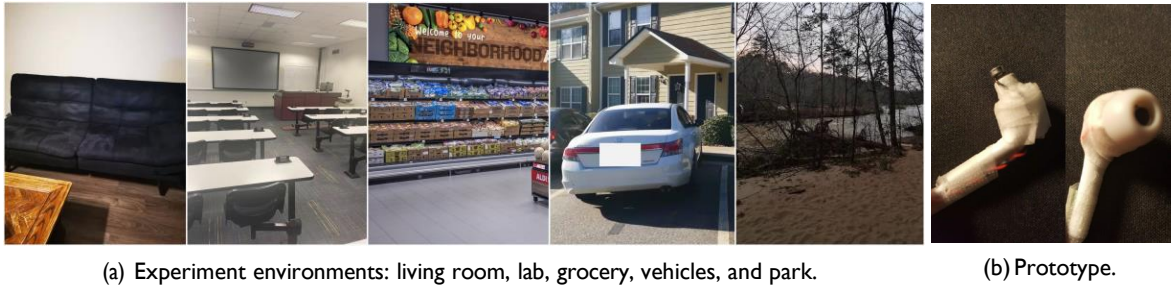

(a) Experiment environments: living room, lab, grocery, vehicles, and park.   (b) Prototype.

Fig. 15. Experiment environment and prototype.

high probability is considered as a sonic event. These sonic events are extracted as our segmented sonic waves. They are then labeled for the next step process.

**Harmonic Portion of Sonic Waves.** Within one segment of sonic waves, to locate when the tapping and sliding occur and also increase SNR, we utilize harmonic ratio to determine the harmonic portion of teeth gestures [31]. For each of our sliding windows, we first determine the normalized autocorrelation of the sonic waves. After that, we estimate the harmonic ratio by finding the maximum of the normalized autocorrelation. And the estimation is then improved using parabolic interpolation. As shown in Fig. 12, the nonharmonic portion with background noises are filtered out. It could be noticed that the harmonic ratio for different gestures also can be utilized to reveal the gaps and spacing between teeth (shown as vertical blue lines in Fig. 12).

**Authentication Models.** We adopt a fully connected neural network as our model for authentication. The toothprint-induced features such as MFCC, pitch, harmonic potion are first standardized and then feed into our training model. The first fully connected layer accepts these features and transforms into each subsequent layer that has a connection to the previous layer. For every fully connected layer, the input is first weighted by optimized parameters and bias vectors. After that, the output of the fully connected layer goes through the activation layer. The activation we choose here is ReLU. Meanwhile, for the last fully connected layer, we utilize softmax activation layer to generated the results of the model and predicted label. With the model and output label, ToothSonic could authenticate legit users and detect the adversary.

## 6 PERFORMANCE EVALUATION

### 6.1 Experimental Setup

**Environments and Hardware.** Our system could be utilized in various environments under everyday use scenarios. People may need authentication for device access, payment, and communication in various places such as living room, lab, grocery, vehicles, and park. These environments are shown in Fig. 15(a). We evaluate our system in these real-world environments. Moreover, we ask the participants to wear our prototype in their natural habits as to how they would wear earables every day to better simulate the real daily usage of ToothSonic. They are also allowed to maintain different postures such as sitting, standing, walking, and driving.

Existing commercial earables equipped with inward-facing microphones are less desirable because their firmware usually setup the restriction from getting access to the raw data. To this end, we built our prototype of earables with only off-the-shelf hardware to demonstrate its practicability and compatibility, as shown in Fig. 15(b). We utilize one microphone chip with a sensitivity of -28±3 dB to implant into commonly seen earbuds with a 3.5 mm audio jack and 12mm speaker. The total cost of the equipment is just several dollars. The earable prototype in our system is more affordable to a wider range of customers compared to the commercial products.

| No. | 1. Teeth Gestures | 2. Comfortable | 3. Less Comfortable | 4. Have Difficulties | 5. Can not perform |
|---|---|---|---|---|---|
| 1 | Occlusion Sliding | 92% | 4% | 4% | 0% |
| 2 | Molar Sliding | 84% | 8% | 8% | 0% |
| 3 | Canine Sliding | 76% | 12% | 12% | 0% |
| 4 | Incisor Sliding F/B | 80% | 8% | 12% | 0% |
| 5 | Incisor Sliding U/D | 92% | 4% | 4% | 0% |
| 6 | Incisor Sliding L/R | 100% | 0% | 0% | 0% |
| 7 | Occlusion Tapping | 100% | 0% | 0% | 0% |
| 8 | Molar Tapping | 84% | 8% | 8% | 0% |
| 9 | Canine Tapping | 76% | 12% | 12% | 0% |
| 10 | Incisor Tapping | 88% | 8% | 4% | 0% |

Fig. 16. Usability of teeth gestures.

**Data Collection.** We recruit 25 participants for the experiments including 10 females and 15 males with an age range from 22 to 36, and the mean and standard deviation for the participants' ages are 28.16 and 4.13, respectively. The participants are informed about the goal of our experiments. All the participants gave their informed consent to be subject to the study.

Each participant is asked to wear the prototype at her/his habitual position. Next, they are instructed to perform these ten teeth gestures several time until they are confident to perform these gestures. After that, each participate repeat each teeth gesture at least 50 times for our evaluation study. We adopt 10-fold cross-validation to evaluate the system performance. During the 10-fold-cross validation, our models are across all participants. The data were randomly shuffled for each user, i.e., for each user, 10% of her/his data will be left out with nonoverlap. Each gesture in the collected data is considered as one authentication attempt. Thus, we have 12,500 authentication attempts in total. For mimic attack, each user act as an adversary to attack other persons with his/her own acoustic toothprint for each gesture. For replay attack, the adversaries first eavesdrop on the sonic wave of the victim's toothprint by recording the sound at victim's mouth. Then, they replay the recorded one to the authentication system to conduct spoofing attacks.

**Learning Model and Metrics.** We used the library of neural network classifiers in MATLAB. Specifically, the learning rate is 0.01, the model utilizes the LBFGS as its loss function minimization technique to minimize the cross-entropy loss. The deep learning model with the softmax layer will provide predict labels for user identification. Besides, in the authentication system, each user has her claim of identity. For instance, User A claims herself as User A(legit user), or User C claims herself as User A(attacker). From the view of the authentication system, it needs to determine if the claim of this specific user is legit. In our work, besides the evaluation of the prediction models, we acquire the evaluation metrics from the level of the authentication system. And the evaluation metrics we adopt are FRR, FAR and BAC to evaluate the performance of our system. False Rejecting Rate (FRR) is the measure of the possibility that the system will incorrectly reject an access attempt by a legitimate user. False Acceptance Rate (FAR) is the measure of the ratio that the system will incorrectly accept an access attempt by a user who is an adversary. Balanced Accuracy (BAC) is used to evaluate imbalanced datasets and it is defined as the average of True Positive Rate and True Negative Rate.

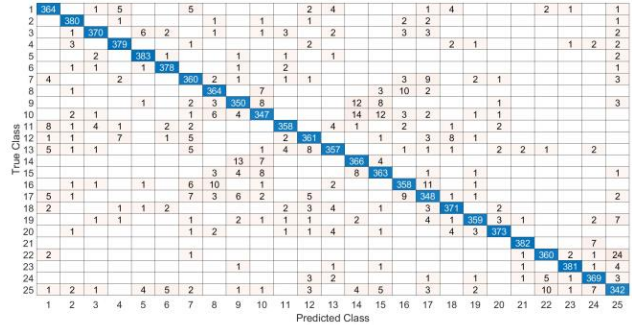

Fig. 17. Confusion matrix of the learning model.

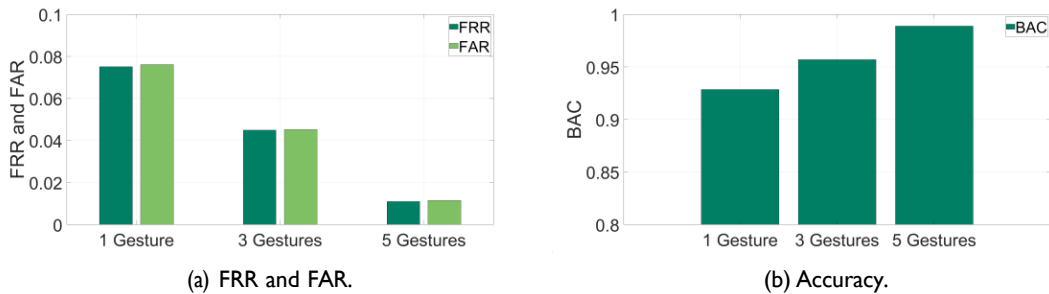

(a) FRR and FAR.

(b) Accuracy.

Fig. 18. Overall performance when using every gesture without considering gesture usability.

### 6.2 Usability of Teeth Gestures

One major factor we considered when we design the teeth gestures is how easily could a user perform the designed gestures. It is thus important to understand the gesture usability from the participants' point of view. We design a survey including three levels of usability, i.e., *comfortable*, *less comfortable*, *have difficulties*, and *can not perform*. The level of *comfortable* means the user could comfortably and consistently perform the gesture, whereas *less comfortable* refer to the cases that the user feel the gesture can be consistently performed after more training efforts. And *have difficulties* means the user feel it is difficult to maintain gesture consistency when performing the gesture multiple times over time. Lastly, the item of *can not perform* means the user found it is challenge to learn the gesture and can't perform the gesture.

Fig. 16 shows the survey results for all the gestures performed by our participates. We can observe that most teeth gestures are very comfortable to perform. For example, all participates prefer *occlusion tapping* and *incisor sliding L/R*. However, we do have a small percentage of participates who have difficulties performing some gestures. For instance, the gestures that involve canine and molar, such as canine tapping and sliding and molar sliding are harder to maintain consistency. For these 10 gestures, the item of can not perform is not selected by all our participants.

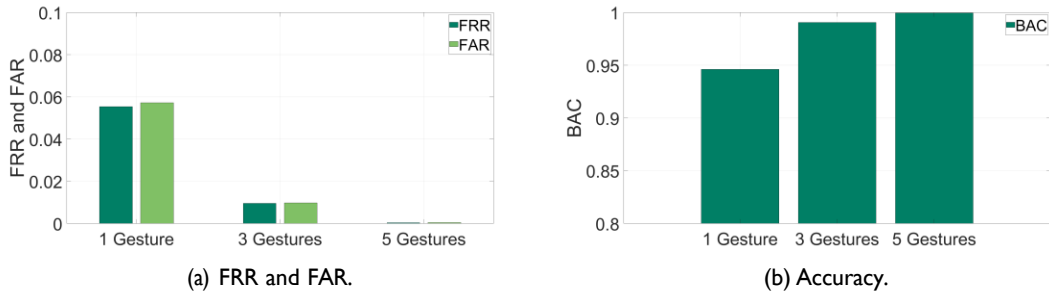

(a) FRR and FAR.   (b) Accuracy.

Fig. 19. Overall performance with users' comfortable gestures.

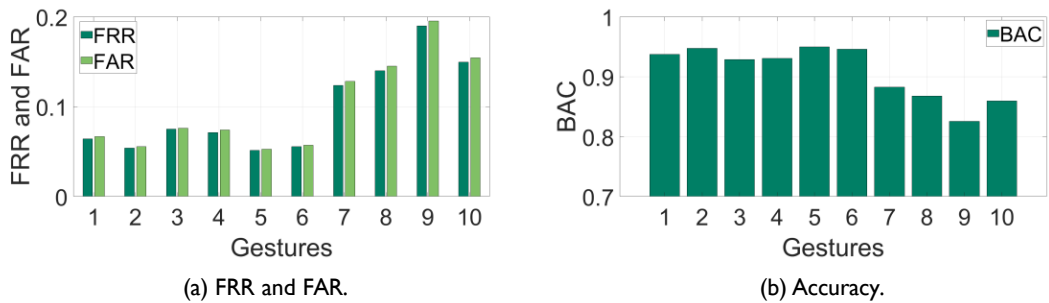

(a) FRR and FAR.   (b) Accuracy.

Fig. 20. Performance for each teeth gesture.

### 6.3 Overall Performance

We study the performance when the participants use different number of gestures for authentication. First, we study the performance of our learning model, the confusion matrix of the learning model when labeling the input biometrics is shown in Fig. 17. Overall, the learning model of our system has a good performance and gives accurate prediction labels based on the input biometrics. Then we evaluate the FRR and FAR from the view of the authentication system and the results are shown in Fig. 18. As shown in Fig. 18(a), after averaging the performance over 1, 3 and 5 gestures, the overall FRR and FAR over 25 subjects for our system is [7.51% 7.61%], [4.49% 4.52%], and [1.10% 1.15%], respectively. We can also find out from the Fig. 18(b), the corresponding accuracy is 92.9%, 95.7% and 98.9%, respectively. These results show that ToothSonic has good performance when using one gesture for authentication. Also, combining multiple gestures is very effective in reducing errors.

We also evaluate the overall performance of our system by using the same comfortable gestures. According to our user survey, the most comfortable gesture is incisor sliding left/right. Fig. 19 shows the results that leverage the same comfortable gestures. As shown in Fig. 19(a), the overall FRR and FAR over 25 subjects for our system is [5.54% 5.72%], [0.96% 0.98%], and [0.04% 0.05%] for 1, 3 and 5 most comfortable teeth gestures, respectively. We can also observe from Fig. 19(b), the corresponding overall accuracy is 94.6%, 99.0% and 99.9%, respectively. These results show that when using the same comfortable gestures, ToothSonic is highly effective in authenticating users. And our system has better performance if gesture usability is considered while combining more gestures is very useful to improve the performance.

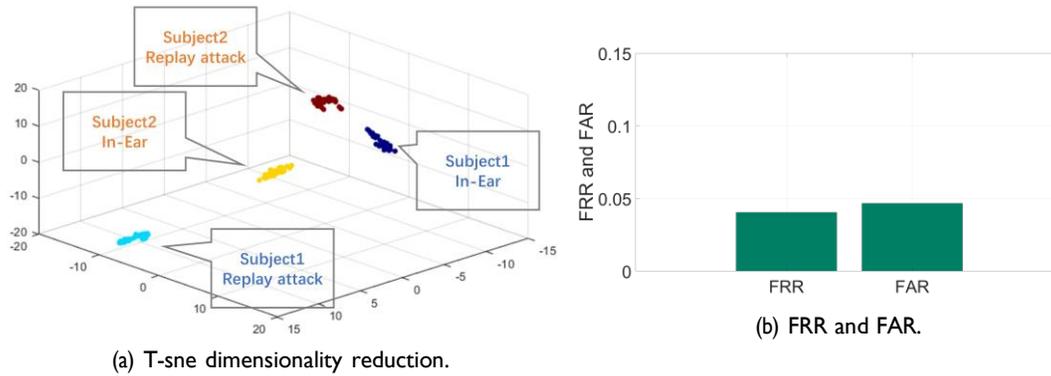

(a) T-sne dimensionality reduction.

(b) FRR and FAR.

Fig. 21. Performance under replay attack.

### 6.4 Impact of Different Teeth Gestures

Fig. 20 shows the performance of each gesture after we average the results from 25 participates. Specifically, No.1 to No.6 stand for the six sliding gestures which are occlusion sliding, molar sliding, canine sliding, incisor sliding F/B, incisor sliding U/D, incisor sliding L/R, whereas the left 4 gestures are tapping gestures including occlusion tapping, molar tapping, canine tapping, and incisor tapping. We can observe that the performance of the sliding gesture is better than that of the tapping gestures. This is because sliding gestures usually have a longer duration and involve more teeth, thus carrying more toothprint information. Therefore, sliding gestures contain more acoustic features than tapping gestures, thus providing better accuracy. Moreover, we find that the gestures with lower usability result in worse accuracy. This is due to the fact that these gestures cannot be performed consistently over time for some users. It implies the gesture consistency or usability significantly affects the system performance. We thus should encourage the users to use their most preferred teeth gestures for authentication.

### 6.5 Replay Attack

The teeth-ear channel acts as a hidden propagation channel to secure the toothprint-induced sonic waves. We are thus interested in how effectively the teeth-ear channel could prevent the replay attack if the adversity replays sonic waves recorded by a microphone that is placed close to the victim's mouth. In replay attacks, the recorded sonic waves propagate directly via the air channel without going through the teeth-ear channel. Fig. 21(a) shows the extracted features in the three-dimensional feature space of two subjects and the corresponding replay attacks. We observe that features of legitimate users and the replay attacks are distinct, indicating the teeth-ear channel effectively secures the sonic waves in a way that can prevent relay attacks. Moreover, Fig. 21(b) shows the performance of our system under replay attacks. The FRR and FAR are 4.05%, and 4.66%, respectively. This result is comparable to the scenarios without replay attacks.

### 6.6 Impact of Different Environments

The need for authentication could happen at various locations. We thus study the performance of our system under different environments with various background noise using one gesture. We choose 5 typical environments: living room, lab with people talking randomly in the background, grocery across the different portions, vehicles, and park. Fig. 22 shows our system performance under these environments. We observe that our system works well across different environments. In particular, for these environments, the FRR is 2.57%, 4.15%, 5.04%, 7.11%, 3.84%, respectively. And the corresponding FAR is 3.24%, 4.22%, 3.83%, 5.55%, 4.63%, respectively. Moreover, we observe that a lower noise level results in a better performance. Specifically, the living room is the quietest environment, thus providing the best performance. These results demonstrate that our system is robust under different noisy environments.

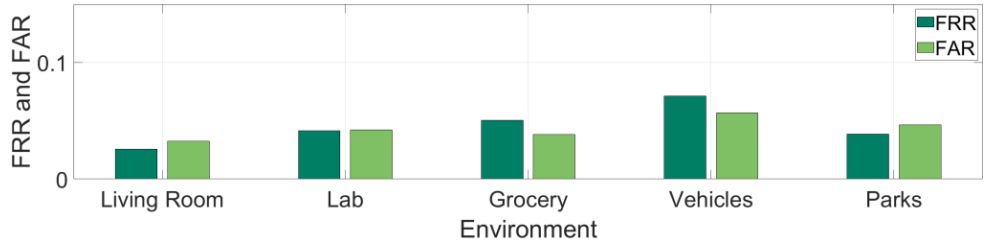

Fig. 22. FRR and FAR under different environment.

### 6.7 Impact of Different Body Motions

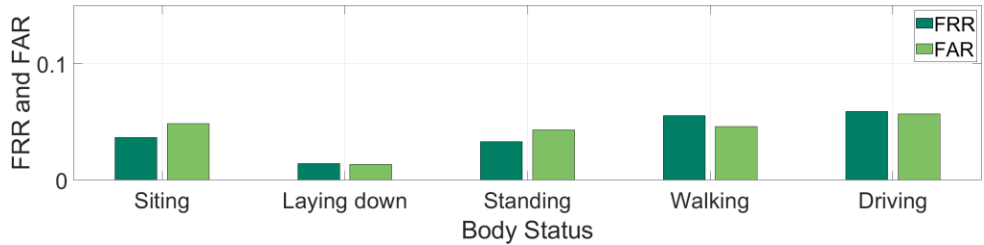

Fig. 23. FRR and FAR across different body motions.

Next, we study how different body motions will affect the performance of our system using one gesture. We consider daily life body motions including sitting, laying down, standing, walking, and driving. To exclude the impact of noise, we collected the data with the engine off for the driving case. As shown in Fig. 23, our system can maintain high accuracy when the users are in different motions. Particularly, for each type of body motion, the FRR is 4.66%, 1.41%, 3.30%, 5.55%, 5.90%, respectively. And the corresponding FAR is 4.85%, 1.37%, 4.30%, 4.58%, 5.69%, respectively. We can observe that the fewer interference vibrations, the higher the accuracy of our system. This is because the body vibrations could be also captured by our earables and interfere with the toothprint-induced sonic waves. Nevertheless, our system still achieves high accuracy under different body motions.

### 6.8 Impact of Different Users

We also study the impact of different users. To study the impact of each user, we remove each user one by one and calculate the rest FAR/FRR. Fig. 24 shows the FRR and FAR after we remove each user. We observe that for some cases, the FAR and FRR are better. After we check out the user survey, we find that these subjects with better results are the ones who have more comfortable gestures to perform. For instance, subjects 11 and 13 have difficulties in both canine tapping and sliding, thus their performance is impacted due to the inconsistency of performing these gestures. However, in general, we could observe that the impact of user differences is restively little.

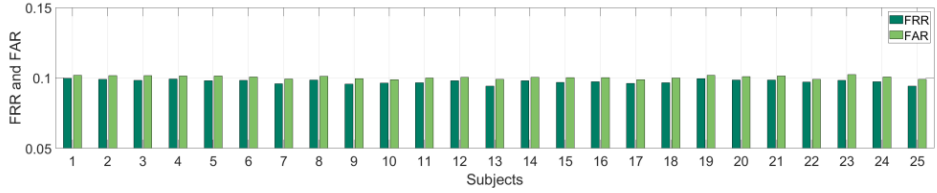

Fig. 24. FRR and FAR across 25 different subjects.

| | Occlusion Sliding | Molar Sliding | Canine Sliding | Incisor Sliding F/B | Incisor Sliding U/D | Incisor Sliding L/R | Occlusion Tapping | Molar Tapping | Canine Tapping | Incisor Tapping |
|---|---|---|---|---|---|---|---|---|---|---|
| Occlusion Sliding | 100.00% | 57.14% | 50.00% | 57.14% | 57.14% | 69.23% | 71.43% | 42.86% | 50.00% | 42.86% |
| Molar Sliding | 57.14% | 100.00% | 54.55% | 50.00% | 38.46% | 38.46% | 42.86% | 60.00% | 41.67% | 23.08% |
| Canine Sliding | 50.00% | 54.55% | 100.00% | 88.89% | 70.00% | 70.00% | 35.71% | 25.00% | 45.45% | 50.00% |
| Incisor Sliding F/B | 57.14% | 50.00% | 88.89% | 100.00% | 80.00% | 80.00% | 42.86% | 23.08% | 41.67% | 60.00% |
| Incisor Sliding U/D | 57.14% | 38.46% | 70.00% | 80.00% | 100.00% | 80.00% | 53.85% | 33.33% | 54.55% | 77.78% |
| Incisor Sliding L/R | 69.23% | 38.46% | 70.00% | 80.00% | 80.00% | 100.00% | 42.86% | 23.08% | 41.67% | 60.00% |
| Occlusion Tapping | 71.43% | 42.86% | 35.71% | 42.86% | 53.85% | 42.86% | 100.00% | 63.64% | 72.73% | 63.64% |
| Molar Tapping | 42.86% | 60.00% | 25.00% | 23.08% | 33.33% | 23.08% | 63.64% | 100.00% | 66.67% | 40.00% |
| Canine Tapping | 50.00% | 41.67% | 45.45% | 41.67% | 54.55% | 41.67% | 72.73% | 66.67% | 100.00% | 66.67% |
| Incisor Tapping | 42.86% | 23.08% | 50.00% | 60.00% | 77.78% | 60.00% | 63.64% | 40.00% | 66.67% | 100.00% |

Fig. 25. Correlation across 10 different gestures.

## 7 LIMITATIONS AND DISCUSSION

**Increase Discriminative Power.** Due to the impact of the COVID-19 pandemic and limited manpower, our evaluation is limited to a small number of participates. While we are unable to recruit a large number of participates to stress-test the system, we can study how the system can effectively support more users if we combine multiple gestures. Of course, combining more gestures could generate more information to represent toothprint, thus increasing the number of users the system can support. When selecting multiple gestures, we should prefer to choose these with small correlations. As if we choose more less-correlated gestures, the discriminative power can increase exponentially compared to that of a single gesture. We thus investigate what is the correlation between gestures based on the overlapped toothprint information carried by the gestures that described in Fig. 8. Fig. 25 illustrates the confusion matrix of the correlation between gestures. In high-level, sliding gestures are less correlated with tapping gestures. We thus should include at least one sliding and one tapping gesture when choosing multiple gestures. Based on the gesture usability, occlusion sliding and tapping are preferred ones. Moreover, occlusion sliding has the lowest correlation with canine sliding, whereas occlusion tapping is less correlated with molar and incisor tapings. Thus, canine sliding, molar, and incisor tapings are the next preferred ones if we want to involve even more gestures. More details could be found in Fig. 25.

**Dental Works and Aging.** Getting dental works after enrollment will affect user authentication. For example, orthodontic treatment will gradually change the dental geometry, and dental fillings add artificiality to teeth. Moreover, some parts of the tooth, such as the enamels, gradually wear as people aging, which may degrade the system accuracy. Therefore, people with dental works or aging persons will need to update their profiles from time to time or an adaptive profile updating method should be deployed [8].

**User Privacy.** While ToothSonic is promising in earable authentication, it could also raise privacy concerns if earables transmit the sensed acoustic toothprint to external devices or servers for identity verification. Transmitting acoustic toothprint increases the attack surface for an adversary to hack the acoustic toothprint. As toothprint is not retractable, one user cannot change or replace it. Leaking the acoustic toothprint could potentially compromise user identity. It is thus more desirable to run the authentication on earables to protect user privacy.

**Earable Hardware.** On-earable implementation requires earables have sufficient computational power to process the acoustic signals in real-time. Although current earables are increasingly equipped with various sensors, a few of them have such a computational capacity. Encouragingly, recent releases of the Apple H1 chip in the Airpods Pro and QCS400 by Qualcomm [10] are capable to support voice-based on-device AI. It implies that implementing ToothSonic on earable could be realized in near future. Moreover, energy consumption could be another issue as both microphone and computing chips on earable operate at a non-trivial energy floor. ToothSonic thus is more suitable for on-demand authentication if energy becomes a bottleneck of earables.

**Sensing While Playing Music.** Further earables may run multiple applications simultaneously. One limitation of the current ToothSonic is that it cannot be used while playing music as the music will interfere with the toothprint-induced sonic waves. One possible way to separate the toothprint-induced sonic waves is to feed the original music sound waves to the earable so as to subtract the music signals from the sensed acoustic signals. It, however, may be more viable to pause the music and switch to authentication mode. The music can resume after user authentication.

**Implicit Authentication.** Functional biometrics, such as toothprint in ToothSonic, could be naturally combined with implicit authentication to enable some emerging applicaiton [25, 35, 60]. Toothprint could be the desired approach for interaction methods in VR and AR applications to enable implicit authentication. When using tooth gestures to input sensitive information, the toothprint could be utilized as input and implicit authentication simultaneously. Nevertheless, other existing authentication and interaction methods became less comfortable to use or more vulnerable to attacks since the view of the user is entirely blocked by the devices. For instance, leveraging hand gestures to input PIN in a virtual keyboard is vulnerable to video-recording attacks. And voices sample could be easily recorded for the mimic attack without user knowledge. We will explore how ToothSonic could be extended to more applications on implicit authentication and human computer interaction in the future.

**Data collection.** The data used in this paper were collected during the impact of the COVID-19 pandemic, thus only a limited number of users participating in our experiments. Experiments with more subjects could better evaluate our system performance. We leave this as one of our future works along with other earables experiments. We will also enhance the usability evaluation of our future participants, for instance, using a 5/6/7-point Likert item for our future survey.

## 8 CONCLUSION

In this work, we propose ToothSonic, a secure earable authentication system that leverages the toothprint-induced sonic waves for user authentication. ToothSonic has several advantages over traditional biometrics, such as anti-spoofing, wide acceptability, and potentially implicit authentication when gestures are also used for a computer interface. Our study shows that the designed teeth gestures could produce effective sonic waves that carry the toothprint information, and the extracted multi-level acoustic features are very effective in authenticating users.

Experiment studies with 25 participants show that ToothSonic achieves up to 95% accuracy with only one of the users' gestures. And combining more gestures could further improve the authentication accuracy.